\documentclass[preprint,aps,pra]{revtex4}

\usepackage{graphicx}
\usepackage{amsfonts}
\usepackage{amssymb}

\usepackage[normalem]{ulem}

\begin{document}

\title{Off--diagonal Ground State Properties of a 1D Gas of Fermi Hard Rods}

\author{F. Mazzanti$^1$, G. E. Astrakharchik$^2$, J. Boronat$^2$,
and J. Casulleras$^2$}

\address{ $1$ Departament de F\'{\i}sica i Enginyeria Nuclear, Comte Urgell
  187, Universitat Polit\`ecnica de Catalunya, E-08036 Barcelona, Spain}
\address{ $2$ Departament de F\'{\i}sica i Enginyeria Nuclear, Campus Nord
  B4-B5, Universitat Polit\`ecnica de Catalunya, E-08034 Barcelona, Spain}


\begin{abstract}
A variational Monte Carlo calculation of the one-body density matrix
and momentum distribution of a system of Fermi hard rods (HR) is
presented and compared with the same quantities for its bosonic
counterpart. The calculation is exact within statistical errors since
we sample the exact ground state wave function, whose analytical
expression is known. The numerical results are in good agreement with
known asymptotic expansions valid for Luttinger liquids. We find that
the difference between the absolute value of the bosonic and fermionic
density matrices becomes marginally small as the density increases. In
this same regime, the corresponding momentum distributions merge into 
a common profile that is independent of the
statistics. Non--analytical contributions to the one--body density
matrix are also discussed and found to be less relevant with
increasing density.
\end{abstract}


\narrowtext

\maketitle

\section{Introduction}

Quantum one--dimensional (1D) systems of bosons and fermions have
attracted great attention in the last years, both from the
experimental and theoretical points of view~\cite{Paredes2004,
Bloch2005, Moritz2003,Richard2003,Khodas2007,Khodas2007b}.  The
combined effect of quantum fluctuations and reduced dimensionality
produces new and intriguing features different or not present in
three-dimensional (3D) systems, as for instance the nonexistence of a
true Bose--Einstein condensate in the homogeneous phase (not even at
$T=0$).

As in the boson case, quasi--1D systems of fermions are experimentally
realized confining to zero point oscillations the radial motion of a
3D trapped cloud of atoms in a largely elongated harmonic trap.  This
is done acting on the system with two perpendicular laser beams
forming a two--dimensional optical lattice.  In this setup, the
transverse frequency $\omega_\bot$ of the harmonic trap is much larger
than the longitudinal frequency $\omega_{\|}$, and such that the Fermi
energy of the confined gas satisfies the condition
$E_F=N\hbar\omega_{\|}/2 \ll \hbar\omega_\bot$. Under these
circumstances, an array of quasi--1D systems is created, thus allowing
for a statistical evaluation of the most relevant quantities of
interest~\cite{Giorgini2007}.

Due to the Pauli exclusion principle, $s$--wave scattering between
atoms of the same spin is not possible.  In this way, the low--energy
properties of the system are dominated by the interaction between
atoms of different spin.  In a pseudopotential description, the
quasi--1D scattering length of the resulting interaction is directly
related to the real 3D scattering length through the expression
\begin{equation}
a_{1D} = {a_\perp^2 \over a_{3D}}\left( 1 - C {a_{3D} \over a_\perp}
\right) \ , 
\label{a1d}
\end{equation}
with $C=\zeta(1/2)/\sqrt{2}=1.0326$ and $\zeta(\cdot)$ the Riemann
  zeta function~\cite{Olshanii1998}.

A peculiarity of the one-dimensional world is that a certain number of
exactly solvable many-body systems is known. The ground state energy
of a system of bosons with $\delta$-pseudopotential interactions
(Lieb-Liniger model) has been obtained \cite{Lieb1963} by using a
Bethe {\it ansatz} solution in all interaction regimes, ranging from
the weakly interacting Bose gas to the strongly correlated
Tonks-Girardeau limit~\cite{Girardeau1960}. The ground state energy of
a system of particles interacting through a $1/z^2$ potential
(Calogero-Sutherland model) is also known for both fermions and
bosons~\cite{Sutherland1971}. Contrary to the case of bosons,
Fermi statistics preclude atoms from interacting via
$\delta$-pseudopotentials, which are commonly used to describe
$s$--wave scattering processes.  Instead, $\delta$-scattering is
permitted between atoms of different spins in a system of two
component fermions. Such systems have attracted great interest for a
long time. There are two classes of excitations in homogeneous
two-component 1D Fermi gases, density and spin modes. In the case of
attractive interactions sound waves propagate with a well defined
velocity while spin waves exhibit a gap~\cite{Krivnov1975}. If the
attraction is very strong, composite molecules consisting of two
fermions of different spin are formed.

Even if the energetic properties of a few 1D systems can be determined
exactly, much less is known about the correlation functions. Some
short-distance properties (at least the leading terms in a $z\to 0$
expansion) are known for the one-~\cite{Olshanii2003},
two-~\cite{Gangardt2003} and three-~\cite{Cheianov2006} particle
correlation functions of the Lieb-Liniger model.  The Tonks-Girardeau
system is a special case where a finite number of terms in the
expansion of the one-body density matrix is enough to provide a
meaningful description of this quantity at all
distances~\cite{Lenard1964}. Moreover, it was noted in
Refs.~\cite{Sutherland1971,Dyson62} that random matrix theory can be
used to describe static density correlators in the Calogero-Sutherland
model for certain values of the interaction parameter ($\lambda=1/2,
2$ and $1$, the later case being equivalent to the Tonks-Girardeau
system).

While the short-range properties of spatial correlators depend
explicitly on the shape of the interaction potential, the long-range
properties are quite generic and governed by the presence of phonons.
The Luttinger model~\cite{Luttinger1963} describes universal
long-range properties of all one-dimensional systems with a phononic
(i.e. {\em linear}) excitation spectrum at low momenta. The long range
description of the main ground state one- and two- particle
correlation functions was obtained by Haldane in
1981~\cite{Haldane1981}. He described asymptotic series with universal
power exponents (i.e. exponents that depend only on the density and
the speed of sound) and non-universal series coefficients.  All terms
of the long-range expansion (with explicit expressions for the
coefficients) of the one-~\cite{Astrakharchik2006} and
two-~\cite{Gangardt2001} particle correlation functions have been
obtained using the replica method for the bosonic and the fermionic
Calogero-Sutherland model, and for all possible interaction strengths
including the Tonks-Girardeau limit.

In spite of the progress achieved in analytical approaches, so far the
only systematic way to obtain a complete description of the
correlation functions relies on numerical methods. Recently, Monte
Carlo methods have been successfully used to obtain these functions
for the Lieb-Liniger~\cite{Astrakharchik2003} and the hard rod
models~\cite{Mazzanti2007,Gregory2005}. Moreover, time-dependent
Lieb-Liniger correlators have also been obtained using numerical
summations of Bethe states~\cite{Caux2006}. In much the same way,
finite temperature correlators for the Lieb-Liniger model have been
calculated using density matrix renormalization
techniques~\cite{Schmidt2007}.

In this article we resort to Monte Carlo methods to analyze the most
relevant ground state correlation functions of a single component
system of fermionic hard rods. We note that while fermions with the
same spin can not interact through a $\delta$-pseudopotential due to
the Pauli exclusion principle, a hard rod interaction is still
permitted and provides the simplest interatomic potential defined by
only one parameter $a$ (the size of the hard-rod which equals its
scattering length).

\section{Results}

In a preceding paper~\cite{Mazzanti2007}, the leading ground state
properties of a Bose gas of 1D hard rods were analyzed and
discussed. In this work we extend this discussion to a population of
fully polarized fermions interacting through the same
spin--independent potential, namely
\begin{equation}
V_{HR}(z) = 
\left\{
\begin{array}{cc}
+\infty & \mbox{   for $\mid\!\!z\!\!\mid \leq a$} \\
0 & \mbox{   otherwise \ ,}
\end{array}
\right.
\label{VHR}
\end{equation}
corresponding to the many--body Hamiltonian
\begin{equation}
H = -{\hbar^2 \over 2m} \sum_{j=1}^N {\partial^2 \over \partial z_j^2}
+ \sum_{i<j} V_{HR}(z_{ij}) \ .
\label{Hamilton}
\end{equation}
Despite the complexity of the interaction, both the ground state wave
function and the energy are exactly known. The former is a Slater
determinant of plane waves~\cite{Nagamiya1940,Krotscheck1999}
\begin{eqnarray}
\Psi_0(z_1, z_2, \ldots, z_N) 
\!\! & = & \!\!
{1\over \sqrt{N!}} 
{\rm det}\left[ {1\over\sqrt{L'}} \exp( i p_k x_k ) \right]
\label{wavefun} \\ [2mm]
& = & \!\!
{2^{N(N-1)/2} \over \sqrt{N!} (L')^{N/2}} 
\prod_{i<j} \sin\left[{\pi\over L'}(x_j-x_i) \right]
\nonumber 
\end{eqnarray}
where $N$ is the total number of particles located in a box of length
$L$ with periodic boundary conditions. In this expression, $L'=L-a N$
is the {\em unexcluded} length, while $p_k'=2\pi n_k/L'$ with $n_k$ an
integer in the range $-N$ to $+N$ plays the role of a
single--quasiparticle momentum. Additionally, $\{x_k=z_k - (k-1) a\}$
are a set of reduced coordinates for a given ordering of the true
particle coordinates $z_1< z_2-a < z_3 - 2a < \cdots < z_N - a(N-1)$.
The ground state energy corresponding to this wave function reads
\begin{equation}
{E_{HR}\over N} = {\pi^2 \hbar^2 n^2 \over 6m} {1 \over (1 - n a)^2} \ ,
\label{EnergyHR}
\end{equation}
with $n=N/L$ the linear density of the system.  As
happens in the 3D case of hard spheres, the scattering length of the
hard rod potential equals the size of the rod, $a_{1D}=a$.

One important consequence of the constrains imposed by the restricted
dimensionality in 1D is the {\em fermionization} of bosons interacting
through diverging potentials, where the strong repulsion between
particles mimics the effect of the Pauli exclusion principle. In the
particular case of hard rods, this duality is explicitly manifested in
the form of $\Psi_0$, as in the Bose case the exact ground state wave
function becomes the absolute value of the expression in
Eq.~(\ref{wavefun})~\cite{Girardeau1960}. In this case, therefore, all
local quantities depending exclusively on $\mid\!\Psi_0\!\mid^2$, such
as the energy, the static structure factor and the two--particle
radial distribution function, are identical for bosons and
fermions. However, other quantities not diagonal in configuration
space are different due to the symmetry properties of $\Psi_0$. In
this work, we analyze two of the most relevant non--diagonal ground
state quantities, namely the one--body density matrix
\begin{equation}
\rho_1(z) = N 
{
\int dz_2 \cdots dz_N 
\Psi_0(z, z_2, \ldots, z_N) \Psi_0(0, z_2, \ldots, z_N)
\over 
\int dz_1 dz_2 \cdots dz_N \Psi_0^2(z_1, z_2, \ldots, z_N) 
} \ .
\label{rho1}
\end{equation}
and its Fourier transform, the momentum distribution
\begin{equation}
n(k) = {1\over 2\pi n} \int dz e^{ikz} \rho_1(z) \ .
\label{momdist}
\end{equation}
Both quantities have been evaluated via Monte Carlo sampling of the
ground state wave function in Eq.~(\ref{wavefun}) for a number of
fermions $N$ between 125 and 1001, located in a box of length $L$ with
periodic boundary conditions. Notice that since we use the analytical
solution corresponding to the ground state of the
Hamiltonian~(\ref{Hamilton}), the results of the simulation are exact
in a statistical sense.

The one--body density matrix satisfies the condition
$\rho_1(0)/n=1$, which is a direct consequence of translational
invariance and normalization. In the absence of a Bose--Einstein
condensate, $\rho_1(z)$ decays to zero at large distances. For a
system of fermions with a non--positive definite wave function,
$\rho_1(z)$ changes sign with increasing distance. This is the case
for instance of the 1D free Fermi gas, where one has in the
thermodynamic limit
\begin{equation}
\rho_{1D}^{FFG}(z) = {\sin(\pi n z)\over\pi z} \ ,
\label{rho1-1DFFG}
\end{equation}
with an infinite number of nodes located at the points $z_m=m/n$ with
$m$ an integer. In the HR system at low densities, the average
distance between particles is much larger than the rod size and the
net effect induced by the potential is equivalent to a point-like
boundary condition. In this limit $\Psi_0$ approaches the ground state
wave function of the 1D free Fermi gas (as can be easily checked from
Eq.~(\ref{wavefun}) setting $a\to 0$), and therefore
$\rho_1(z)\approx\rho_{1D}^{FFG}(z)$. 

\begin{figure}[b!]
\begin{center}
\includegraphics*[width=0.45\textwidth]{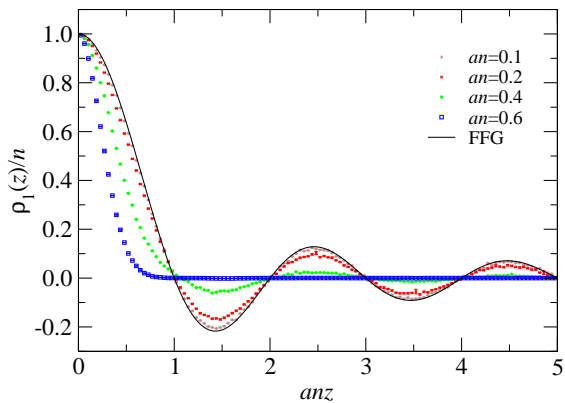}
\end{center}
\caption{(Color online) $\rho_1(z)$ at the rod
  densities $n a=0.1$, $0.2$, $0.4$ and $0.6$ (blue pluses, red
  crosses, green stars
  and blue squares) compared with the one--body density matrix of the 1D
  free Fermi gas (solid line.)}
\label{fig-onebody-1}
\end{figure}

The one--body density matrix $\rho_1(z)$ at the particle densities
$an=0.1, 0.2, 0.4$ and $0.6$ (pluses, crosses, stars and squares) is
compared with $\rho_{1D}^{FFG}(z)$ (solid line) in
Fig.~\ref{fig-onebody-1}. As it can be seen, little differences
between $\rho_1(z)$ and $\rho_{1D}^{FFG}(z)$ arise at densities lower
than $an\approx 0.1$. As the density increases, the main structure of
the low--density $\rho_1(z)$ is kept while the strength at every point
is depressed compared with $\rho_{1D}^{FFG}(z)$. At intermediate and
high densities $na \gtrsim 0.5$ the oscillations are no longer visible
at the scale of Fig.~\ref{fig-onebody-1}, and essentially all of the
strength of $\rho_1(z)$ is located around the origin. It is also worth
to notice from the figure that the nodal structure of the one--body
density matrix is poorly affected by particle correlations, keeping
the nodes of $\rho_1(z)$ remarkably close to those of
$\rho_{1D}^{FFG}(z)$.

The one--body density matrix of any fermionic Luttinger liquid (as
hard rods) admits the following asymptotic expansion valid at large
distances $z\gg n^{-1}$~\cite{Haldane1981}
\begin{equation}
{\rho_1(z) \over n} = 
{1\over(n\mid\!\!z\!\!\mid)^{1/\eta}} \sum_{m=0}^\infty
{F_m 
\sin\left[2\pi n\left(m+{1\over 2}\right)\!\mid\!\!z\!\!\mid\right]
\over (n\mid\!\!z\!\!\mid)^{(m+{1\over
      2})^2\eta}}\,
\label{haldane-b1}
\end{equation}
where $\eta=2K$ and $K=\pi\hbar n/m c$ is the Luttinger parameter
written in terms of the sound velocity $c$.  The coefficients $F_m$
and the value of $\eta$ change with the interaction and are therefore
model--dependent. For a free Fermi gas $\eta=2$, $F_0=1/\pi$ and
$F_{m>0}=0$. Apart from a few remarkable
cases~\cite{Lenard1964,Astrakharchik2006}, these coefficients are, in
general, unknown.  For a system of hard rods, bosons or fermions,
$\eta=2(1-an)^2$ which is positive and decreases from $2$ to $0$ with
increasing density $an\in[0,1]$.  High order terms in
Eq.~(\ref{haldane-b1}) turn out to be then less relevant at large
distances, and the asymptotic behavior of $\rho_1(z)$ when
$z\to\infty$ is therefore dominated by the $m=0$ term.  This term
cancels at the nodes of $\rho_{1D}^{FFG}(z)$, that is, at the
positions $\{ z_m =m/n \}$. At these points all other terms in the
series vanish too, and the whole expression is zero. This fact
explains why the nodes of $\rho_1(z)$ and $\rho_{1D}^{FFG}(z)$ are so
close.  Still, Eq.~(\ref{haldane-b1}) is an asymptotic expansion valid
only beyond some healing distance, and therefore the first nodes of
$\rho_{1D}(z)$ can deviate from those of $\rho_{1D}^{FFG}(z)$. This
effect is small even at large densities and can hardly be appreciated
in the $a n=0.4$ case of Fig.~\ref{fig-onebody-1}.

\begin{figure}[t!]
\begin{center}
\includegraphics*[width=0.45\textwidth]{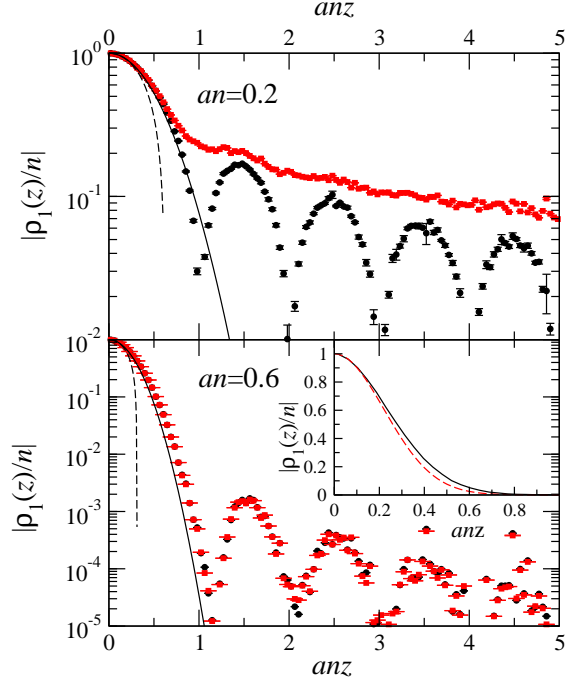}
\end{center}
\caption{(Color online) Absolute value of the one-body density matrix
  in logarithmic scale at the particle densities $n a=0.2$ and
  $an=0.6$ (upper and lower panel, respectively). Black circles stand
  for fermions, red squares for bosons. The dashed and solid lines
  represent the approximations reported in Eqs.~(\ref{rho1lowx})
  and~(\ref{cumul_rho1}). Inset: $\rho_1(z)/n$ compared with the
  truncated cumulant expansion of Eq.~(\ref{cumul_rho1}) (solid and
  dashed lines, respectively).}
\label{fig-onebodyFB-1}
\end{figure}

Another relevant aspect concerning the structure of the one--body
density matrix of a fermionic system of hard rods is manifested when
$\rho_1(z)$ is compared with $\rho_{1B}(z)$, the one--body density
matrix of a system of boson hard rods of the same length and mass, and
at the same density. Notice that $\rho_{1B}(z)$ is built as in
Eq.~(\ref{rho1}) but starting from the corresponding bosonic hard rod
ground state wave function, which is nothing but the absolute value of
the wave function $\Psi_0(z_1, z_2, \ldots, z_N)$ of
Eq.~(\ref{wavefun})~\cite{Mazzanti2007}.  Figure~\ref{fig-onebodyFB-1}
displays the absolute value of $\rho_1(z)$ compared with
$\rho_{1B}(z)$ (which is positive definite) for two particle
densities $an=0.2$ and $an=0.6$ (upper and lower panels,
respectively). As it can be seen from the figure, both functions share
a common short-distance behavior. It is easy to see from the symmetry
of the one--body density matrix and the definition of the momentum
distribution in Eq.~(\ref{momdist}) that the leading $z\to 0$
behavior of $\rho_1(z)$ and $\rho_{1B}(z)$ is equal and given by
\begin{equation}
{\rho_1(z)\over n} = 1 - {\pi^2 n^2 \over 6(1-an)^2} z^2 + {\mathcal
O}(z^4) \ ,
\label{rho1lowx}
\end{equation}
where the coefficient of $z^2$ is proportional to the kinetic energy
per particle which, for a system of hard rods (bosons or fermions),
equals the total energy in Eq.~(\ref{EnergyHR}). The prediction of
Eq.~(\ref{rho1lowx}) is shown as a dashed line in
Fig.~\ref{fig-onebodyFB-1} for the two densities analyzed. Clearly,
the quadratic approximation is too crude to describe $\rho_1(z)$ but
at very short distances. A better approximation is obtained when a
truncated cumulant expansion is considered instead
\begin{equation}
{\rho_1(z) \over n}  \approx \exp \left[ 
-{\pi^2 n^2 \over 6 (1-an)^2}\,z^2 \right] \ ,
\label{cumul_rho1}
\end{equation}
which is also displayed in the upper and lower panels of
Fig.~\ref{fig-onebodyFB-1} with a solid line.  As can be seen from the
figure, the cumulant approximation works well both at low and high
densities up to approximately the first node of $\rho_1(z)$. Beyond
that point the approximation certainly breaks down because cumulant
expansions can only be carried out on positive defined functions while
$\rho_1(z)$ changes sign. In this sense, Eq.~(\ref{cumul_rho1}) is a
better approximation to $\rho_{1B}(z)$, although visible differences
remain in the tails when plotted in logarithmic scale. At intermediate
and high densities, however, there is no practical distinction between
the fermionic and bosonic cases since the strength in $\rho_1(z)$
after the first oscillation is remarkably low. This is seen also in
the inset of Fig.~\ref{fig-onebodyFB-1}, where the dashed line is the
cumulant approximation while $\rho_1(z)$ and $\rho_{1B}(z)$ are
displayed with a solid line and can not be distinguished from each
other.

At distances larger than the position of the first node, $\rho_1(z)$
and $\rho_{1B}(z)$ differ more significantly, specially at low
densities. However, at large densities $\mid\!\rho_1(z)\!\!\mid$ and
$\rho_{1B}(z)$ are almost identical as seen from the figure, the main
difference between the two functions being that $\rho_{1B}(z)$ is
always positive while $\rho_1(z)$ changes sign each time a node is
crossed.

This striking fact can be understood by direct inspection of the
particle configurations contributing to the one--body density matrix.
By definition, $\rho_1(z)$ is related to the probability of destroying
a particle at the origin and creating a new one at a distance $z$, as
given by $\langle \Psi^\dagger(z) \Psi(0)\rangle$ with $\Psi$ and
$\Psi^\dagger$ field operators. In first quantization, $\rho_1(z)$
measures the overlap between the wave functions corresponding to
different particle configurations where one particle has shifted its
position by an amount $z$ (see Eq.~(\ref{rho1})) while all other
particles are kept in their original positions. Clearly, the different
symmetry properties of the Fermi and Bose wave functions can make the
contribution of these configurations to the one--body density matrix
be of different sign.

A remarkable property of 1D systems like the one analyzed in this work
is the fact that the different particle configurations can be
classified in $N!$ disjoint subspaces, according to the ordering of
the particles. In this way, one can attach labels 1 to $N$ to the
particles of a given configuration, and build all other subspaces by
sorting the particles in a different order. One can then move from the
original subspace $\Omega_{0}$ to any other subspace $\Omega_{\alpha}$
by transposition of particle coordinates. In the Bose case, the wave
function is always positive. In the Fermi case, however, moving from
one subspace to another implies a change in sign equal to $(-1)^{P}$,
where $P$ is the parity of the permutation that leads to that subspace
starting from the coordinate ordering of $\Omega_{0}$.

\begin{figure}[t!]
\begin{center}
\includegraphics*[width=0.3\textwidth]{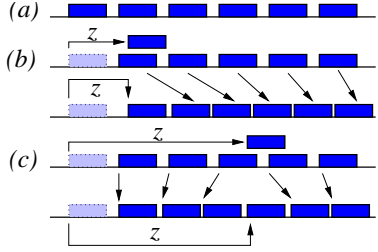}
\end{center}
\caption{(Color online) Three different configurations contributing to
  $\rho_1(z)$. (a) shows the most probable configuration where all
  particles are equally spaced. In (b) one particle is displaced a
  distance $z$, but there is not enough room for it to be placed
  between two other rods. The only configurations in $\Omega_0$ that
  allow this are ones were many particles have been shifted to close
  positions, and these are much less probable at high densities. In
  (c) the displacement $z$ is large and that can only be done moving
  from $\Omega_0$ to another $\Omega_\alpha$, that is, changing the
  order of the particles with respect to the original ordering.}
\label{Fig-moveRods}
\end{figure}

Now consider the high density limit for a system of hard rods. 
Given the wave function~(\ref{wavefun}), the most probable
configuration is one where all particles are equally spaced.
Each rod of size $a$ has, in average, a space $l=L/N$ available.
At high densities, $l$ is only slightly larger than $a$ and there is
not enough room for a particle to be placed between two other
particles. In this sense, the most probable configuration will not
contribute to $\rho_1(z)$ when $z$ is larger than $l$, as
schematically represented in Fig.~\ref{Fig-moveRods}a. The only
configurations in subspace $\Omega_0$ contributing to $\rho_1(z)$ are
those less probable were particles are more packed as in
Fig.~\ref{Fig-moveRods}b, making room for the movement of the selected
particle. Belonging to $\Omega_0$, these configurations have always
positive sign since no particle reordering is required, and contribute
the same for bosons and fermions. Of course, this mechanism is only
possible at short distances and that explains why the low $z$ behavior
of $\rho_1(z)$ and $\rho_{1B}(z)$ is equal.

As the distance $z$ increases, the contribution from $\Omega_0$ to
$\rho_1(z)$ becomes negligible and the main contributions come from
movements involving a change from $\Omega_0$ to different subspaces
$\Omega_\alpha$. As before, and depending on the density, it may
happen that the most probable configuration can not contribute due to
the lack of available space, but there are configurations in other
subspaces were only the movement of a few particles with respect to
the equally spaced configuration is involved, as shown in
Fig~\ref{Fig-moveRods}c.  These are the most probable ones and produce
the major contributions to the one--body density matrix.  These
contributions have a sign $(-1)^P$ for fermions and $+1$ for bosons.
At high densities, only one subspace $\Omega_\alpha$ contribute
significantly, and therefore all contributions have the same sign.  If
this sign is negative, that leads to a total contribution to the
one--body density matrix that is positive for bosons and negative for
fermions but equal in absolute value. That explains the observed
behavior shown in the lower panel of Fig.~\ref{fig-onebodyFB-1}.

It is also clear from the above arguments that this phenomenon does
not happen at low densities, since in this case many different
particle configurations have non-negligible contributions for a given
$z$, and even at large distances configurations from different
$\Omega_\alpha$ subspaces contribute. In the Bose case all these
configurations have the same sign, while in the Fermi one some are
positive and others are negative, leading to a total contribution that
is noticeably different for bosons and fermions.

The form of the exact wave function~(\ref{wavefun}) is quite peculiar
as the long-range decay of the two-body term $\sin (\pi x /L')$ is
very ``weak''. In this way, correlations between particles are
``strong'' even at large distances of the order of the box size. This
situation should be contrasted with two- or three- dimensional
systems, where the two-body Jastrow factor is significantly different
from 1 only at short distances~\cite{Reatto1967}.  As a consequence of
strong correlations, density matrix $\rho_1(z)$ of the HR system does
not saturate to a constant value, but decreases as $z$ increases. This
is the reason of the absence of a true BEC in the one-dimensional
system of hard rods, even at zero temperature.  Furthermore, it is
also easy to understand that the decay is greatly enhanced in the
high-density regime. Indeed, as we see, contributions to the
off-diagonal $|z|\gg n^{-1}$ elements of the one-body density matrix
at high density come mainly from the scenario described in
Fig.~3c. The higher the density, the higher the cost of moving
particles in order to make enough room for the displaced particle, and
the smaller the weight of such a contribution. As a result $\rho_1(z)$
decays faster at higher densities.

Being the one--body density matrix an even function of its argument,
the momentum distribution (its Fourier transform, see
Eq.~(\ref{momdist})) is also even $n(-k)=n(k)$. Consequently and as
originally pointed out by Sutherland~\cite{Sutherland1992}, all odd
moments of $n(k)$ are zero. Moreover, $\langle k^{2n}\rangle$ can be
written in configuration space as the expectation value of
$\langle\sum_j\!\!\mid\!\!\partial^n\Psi/\partial
z_j^n\!\!\!\mid^2\rangle$ once an integration by parts is carried out,
and this is independent of the sign of the wave
function. Consequently, all finite moments of $n(k)$ are equal for
Fermi and Bose hard rods. If $\rho_1(z)$ and $\rho_{1B}(z)$ were
analytic functions of their arguments and the corresponding momentum
distributions had non--divergent moments $\langle k^{2m}\rangle <
+\infty$ for all integer $m\geq 0$, then $\rho_1(z)$ and
$\rho_{1B}(z)$ would be equal, which is not the case.  This can happen
if only a finite number of moments of the momentum distributions
exist, or if the one--body density matrices have non--analytic
contributions. In the case of hard rods, both things happen. On one
side the interatomic potential is a source of non--analyticity to the
wave function, due to the excluded length which makes $\Psi_0(z_1,
\ldots, z_N)$ and all its derivatives be zero when two or more rods
overlap, both for fermions and bosons. We thus see that for fixed
coordinates $z_2$ to $z_N$, $f_{z_2,\cdots, z_N}(z_1)=\Psi_0(z_1, z_2,
\ldots, z_N)$ is a non--analytic function of $z_1$, and one integrates
$f_{z_2, \cdots, z_N}(z)$ to get the one-body density matrix. 

Notice the non--analyticity of $\rho_1(z)$ is an effect produced by
the interatomic potential, and thus it is expected to apply both to
the Bose and to the Fermi cases.  In the $an\to 0$ limit the potential
reduces to a point-like interaction with well known properties. For
bosons, the system enters the Tonks--Girardeau
regime~\cite{Girardeau1960}. For fermions, the effect of the potential
is already taken into account by the antisymmetry of the wave
function, and the system behaves as a 1D free Fermi gas. In the later
case, the corresponding momentum distribution is
$n_{1D}^{FFG}(k)=\theta(k_F-k)$ and all moments exist. On the
contrary, in the Tonks--Girardeau limit $n(k)$ presents a
$1/\!\!\mid\!\!k\!\!\mid^4$ large-$k$ tail that makes all moments
$\langle k^m \rangle$ with $m>3$ diverge~\cite{Olshanii2003}.  The
same conclusions about the relevance of the statistics can be drawn
when the 1D free Bose gas, with $n(k)\propto\delta(k)$ and all moments
equal to 0, is compared with its fermionic counterpart (Fermionic
Tonks Girardeau), which has a momentum distribution
$n(k)=1/[1+(k/2)^2]$~\cite{Girardeau2006}.  Furthermore, the behavior
of the momentum distribution of other exactly solvable 1D models
describing interacting Luttinger liquids can be related to the hard
rod system though the value of the Luttinger parameter. For instance,
in the Calogero--Sutherland model~\cite{Sutherland1971} with an
interatomic potential of the form
\begin{equation}
V(z)={ \lambda(\lambda-1)\pi^2 \over L^2\sin^2(\pi z/L)} \ ,
\label{CalogeroSutherland}
\end{equation}
the momentum distribution for bosons and fermions decays to zero at
large momentum according to a $1/\!\mid\!k\!\mid^{2+2\lambda}$
law~\cite{Astrakharchik2006}, and thus only a finite number of moments
of $n(k)$ exist. Apparently, therefore, the inclusion of an
interatomic potential in a 1D system leads to a power-law decay in
$n(k)$ that makes it possible for the bosonic and the fermionic
systems to have different momentum distributions while sharing all
existing moments. Unfortunately, checking this feature from a Monte
Carlo simulation is very difficult due to the enhancement of the
statistical noise in the tail of $n(k)$ when the order $m$ of the
moment $\langle k^m \rangle$ increases.

\begin{figure}[t!]
\begin{center}
\includegraphics*[width=0.45\textwidth]{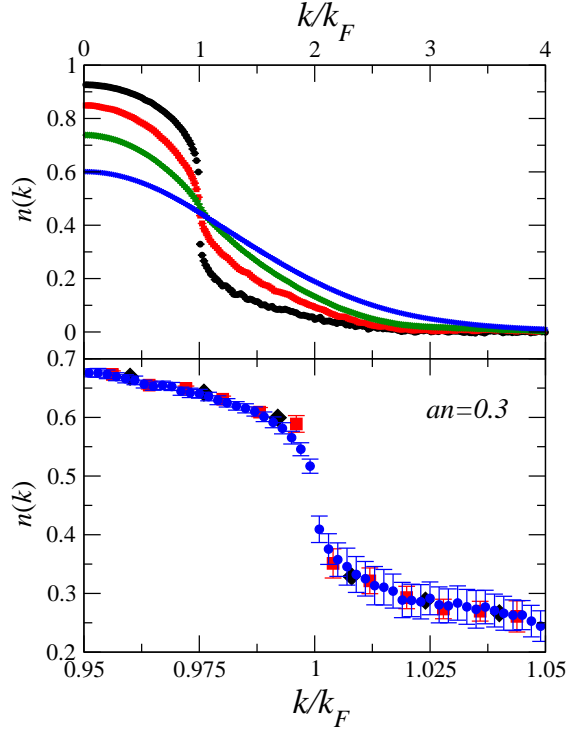}
\end{center}
\caption{(Color online) Momentum distribution of the gas of Fermi hard
  rod at different densities (upper panel). From top to bottom at low
  $k$: $an=0.3$ (black circles), $an=0.4$ (red squares), $an=0.5$
  (green diamonds) and $an=0.6$ (blue stars). The lower panel
  displays $n(k)$ at $na=0.3$ for different number of
  particles: $N=125$ (black diamonds), $N=251$ (red squares), and
  $N=1001$ (blue circles).}
\label{Fig-nk}
\end{figure}

In higher dimensions, the zero temperature momentum distribution
$n(k)$ of a normal Fermi liquid presents a gap at $|k|=k_F$ that is
inversely proportional to the effective mass at the Fermi
surface~\cite{Pines66}. This gap is also present in the 1D Free Fermi
gas. In 1D, however, even weak particle correlations break down this
picture. This is the case of Luttinger liquids~\cite{Luttinger1963}
where Haldane theory applies. The Fourier transform of the leading
($m=0$) term in the asymptotic expansion reported in
Eq.~(\ref{haldane-b1}), which oscillates with frequency $k_F=\pi n$,
reveals that $n(k)$ is continuous at $k=\pm k_F$ when
$1/\eta+\eta/4\geq 1$. For hard rods, where $\eta=2(1-an)^2$ and
$an\in [0,1]$, this condition is always met. In this way, the momentum
distribution of the hard rod system of fermions is continuous at
$k=\pm k_F$ at all densities. Furthermore, the same analysis indicates
that the slope of $n(k)$ at these points is infinite and negative when
$1/\eta+\eta/4<2$. Once again, for hard rods this condition implies
the existence of a threshold density $n_c a=1-\sqrt{2-\sqrt{3}}\approx
0.48$ below which $d n(k)/dk$ at $|k|=k_F$ diverges.

\begin{figure}[t!]
\begin{center}
\includegraphics*[width=0.45\textwidth]{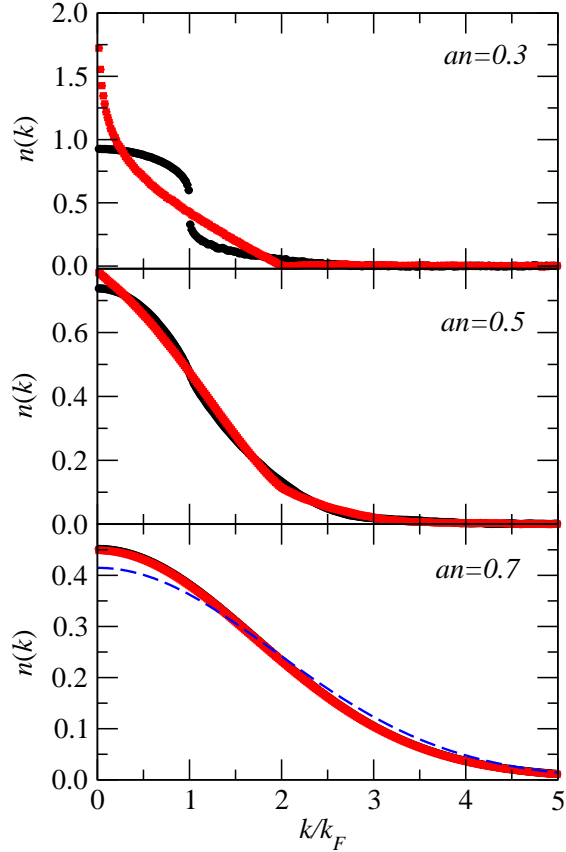}
\end{center}
\caption{(Color online) Momentum distribution of the gas of Fermi
  (black squares) and Bose (red circles) hard rod at different
  densities. The blue dashed line in the lower panel shows the
  momentum distribution associated to the gaussian one--body density
  matrix of Eq.~(\ref{cumul_rho1}).}
\label{Fig-nkFBx}
\end{figure}

These two aspects are illustrated in the upper and lower panels of
Fig.~\ref{Fig-nk}. The upper plot shows the momentum distribution
$n(k)$ of the system at the particle densities $an=0.3, 0.4, 0.5$ and
$0.6$. As can be seen, the behavior of $n(k)$ at $|k|=k_F$ changes
with the density, with steeper slope as the density decreases. These
results are compatible with an infinite derivative at $|k|=k_F$ for
the densities $an=0.4$ and $an=0.3$. In any case, the present
simulations have been carried out for an even number of particles in
order to prevent the ground state from being degenerate.  Since
$k_F=\pi n=\pi N/L$, with a momentum spacing $\Delta k=2\pi/L$ imposed
by the use of periodic boundary conditions, no point falls exactly at
$|k|=k_F$. As a consequence, one can only investigate the behavior of
the momentum distribution at the Fermi momentum by increasing the
number of particles in the simulation.  The lower plot in
Fig.~\ref{Fig-nk} shows $n(k)$ around $k_F$ for $an=0.3$, as a
function of the number of particles in the simulation, for the three
values $N=125, 251$ and $1001$.  As it can be seen from the figure,
the points closer to $k_F$ corresponding to the highest $N$
seem to confirm the theoretical prediction of a continuous $n(k)$ at
that point, while the results obtained at lower $N$, with a coarser
spacing, could lead to the wrong conclusion that there is a gap in
$n(k)$ at $k=\pm k_F$.

We conclude the results section comparing the momentum distribution of
the Fermi and Bose systems of hard rods as a function of the
density. Figure~\ref{Fig-nkFBx} shows $n(k)$ at the rod densities
$an=0.3, 0.5$ and $0.7$ (upper, middle and lower panels,
respectively). As expected, the momentum distributions are quite
different at low densities, as in this limit the fermionic system
approaches the 1D free Fermi gas prediction while the bosonic one
reproduces the Tonks--Girardeau limit. As the density
increases, however, both functions are smeared out and the differences
become less relevant, to the point that both curves overlap and can
not be distinguished from each other at the highest density considered
$an=0.7$.  We have also checked numerically that the lowest order
moments $\langle k^0 \rangle$ and $\langle k^2 \rangle$, the former
being a normalization factor and the later the kinetic energy per
particle (which equals the total energy per particle for hard rods),
are the same for bosons and fermions at the three densities reported
in the figure. Additionally and for $an=0.7$, Fig.~\ref{Fig-nkFBx}
shows with a dotted line the Fourier transform of the gaussian
one--body density matrix of Eq.~(\ref{cumul_rho1}), which is also
gaussian.

It is clear from Fig.~\ref{Fig-nkFBx} that the low density momentum
distributions of both bosons and fermions are far from being gaussian,
but also that the differences between $n(k)$ and $n_B(k)$ reduce when
the density increases. Furthermore, $n(k)$ and $n_B(k)$ seem to
approach a common profile that looks gaussian at low momenta, but this
happens at even higher densities. This means that in the high density
limit the contribution of the non--analytical parts of $\rho_1(x)$ is
less relevant, and also that in this limit the coefficients in the
Taylor expansion of the analytical part of $\rho_1(x)$ are mostly
dominated by the kinetic energy per particle, according to the
cumulant expansion of Eq.~(\ref{cumul_rho1}). It is also apparent from
the figure that the differences between the statistics vanish, as a
function of the density, before the common profile is
approached. Still, one can not deduce from the simulation the
analytical form of the large-$k$ behavior of the momentum
distributions, and in particular whether a true gaussian is reached
when $an\to 1$.

\section{Summary and Conclusions}

In summary, we have studied the one--body density matrix and momentum
distribution of a Fermi gas of hard rods, comparing them to their
bosonic counterparts. We find that $\rho_1(z)$ has a nodal structure
quite close to that of the 1D free Fermi gas, in agreement with
Haldane's theory of Luttinger liquids. We have also discussed the
analytical properties of $\rho_1(z)$ to find that non--analytical
parts become less relevant as the density increases. Our numerical
simulations confirm that the momentum distribution does not present a
sharp gap at the Fermi surface, but has infinite derivative at $k=\pm
k_F$ when the density is lower than a critical value $a
n_c=1-\sqrt{2-\sqrt{3}} \approx 0.48$. Furthermore, $n(k)$ for
fermions and bosons share the same lower order moments while showing
an overall fairly different shape. These differences reduce when the
density increases, and remarkably the Bose and Fermi momentum
distributions approach a common limit at high densities.

\vfill

\begin{acknowledgments}

This work has been partially supported by Grants No.~FIS2005-03142
and~FIS2005-04181 from DGI (Spain), and Grants No.~2005SGR-00343
and~2005SGR-00779 from the Generalitat de Catalunya.

\end{acknowledgments}


\end{document}